\documentclass[10pt,showpacs,letterpaper,runinaddress,prl,twocolumn,lengthcheck,preprint]{revtex4}
\usepackage{amsfonts}
\usepackage{amsmath}
\usepackage{amssymb}
\usepackage{graphicx}

\begin{document}

\title{Cavitation from bulk viscosity in neutron stars and quark stars}
\author{Jes Madsen}
\affiliation{Department of Physics and Astronomy, Aarhus University, DK-8000 \AA rhus C,
Denmark}
\pacs{97.60.Jd, 12.38.Mh, 21.65.Qr, 26.60.Kp}

\begin{abstract}
The bulk viscosity in quark matter is sufficiently high to reduce the
effective pressure below the corresponding vapor pressure during density
perturbations in neutron stars and strange stars. This leads to mehanical
instability where the quark matter breaks apart into fragments comparable to
cavitation scenarios discussed for ultra-relativistic heavy-ion collisions.
Similar phenomena may take place in kaon-condensed stellar cores. Possible
applications to compact star phenomenology include a new mechanism for
damping oscillations and instabilities, triggering of phase transitions,
changes in gravitational wave signatures of binary star inspiral, and
astrophysical formation of strangelets. At a more fundamental level it
points to the possible inadequacy of a hydrodynamical treatment of these
processes in compact stars.
\end{abstract}

\date{September 30, 2009}
\maketitle

It is well-known that the bulk viscosity in dense nuclear matter containing
hyperons as well as in quark matter can be very high and therefore may play
an important role in the damping of compact star oscillations and
instabilities \cite%
{Wang:1985tg,Sawyer:1989uy,Madsen:1992sx,Madsen:1998qb,Madsen:1999ci,Jones:2001ya,Lindblom:2001hd,Nayyar:2005th,Alford:2006gy,Sa'd:2006qv,Sa'd:2007ud,Chatterjee:2007qs,Alford:2007rw,Chatterjee:2007iw,Manuel:2007pz,Alford:2008pb}%
. But so far it has not been realized that bulk viscosity may in fact be so
high that a hydrodynamical treatment of the oscillations and instabilities
becomes unphysical because the effective pressure can become lower than the
corresponding vapor pressure and cause the fluid to break up into fragments.
This may introduce interesting new phenomenology and at least call for a
re-evaluation of various processes in compact stars. The effect is the high
chemical potential, low temperature analogy of the recent demonstration at
low chemical potential and high temperature that an expanding quark-gluon
plasma formed in ultra-relativistic heavy-ion collisions may be mechanically
unstable and cavitate, i.e. break apart into droplets \cite%
{Torrieri:2007fb,Torrieri:2008ip,Rajagopal:2009yw,Song:2009rh}. In the
following cavitation and mechanical instability will be used as synonymous
with such a process, and the term vapor pressure $P_{\text{vap}}$ is
borrowed from ordinary fluid theory to denote the pressure below which the
fluid can cavitate.

The term effective pressure $P_{\text{eff}}$ will be used for the
space-space components of the energy-momentum tensor, which enter the
relativistic hydrodynamical equations,%
\begin{equation}
P_{\text{eff}}=P-\zeta \Theta 
\end{equation}%
where $P$ is the normal (thermodynamic) fluid pressure, $\zeta $ is the bulk
viscosity, and $\Theta \equiv \partial _{\mu }u^{\mu }$ with fluid
four-velocity $u^{\mu }$ is the expansion scalar, which has dimension of
1/time and can be thought of as the rate of expansion (see below). Only
homogeneous and isotropic perturbations will be discussed. Cavitation will
happen when $P_{\text{eff}}$ becomes smaller than the corresponding vapor
pressure, $P_{\text{vap}}$. In the case of heavy-ion collisions or
self-bound strange quark matter $P_{\text{vap}}$ is the vacuum pressure,
i.e. $P_{\text{vap}}=0$, but in the case of a phase boundary inside a
compact star $P_{\text{vap}}=P_{\text{ph}}$, i.e. the pressure at the phase
boundary. Terms including shear viscosity are neglected; such terms could
further reduce $P_{\text{eff}}$, but in most parameter ranges of interest
for cavitation in compact stars the effects of bulk viscosity will dominate.
Only first-order (Navier-Stokes) contributions to pressure are considered
since it has been found that inclusion of second-order terms do not
introduce major changes at least in the context of heavy-ion physics \cite%
{Rajagopal:2009yw}. The bulk viscosity calculated from lattice QCD at the
high temperatures and low chemical potentials relevant for heavy-ion
collisions is at the border-line of being large enough to drive the
effective pressure negative on the short (strong interaction) time-scales
involved \cite{Torrieri:2007fb,Torrieri:2008ip,Rajagopal:2009yw,Song:2009rh}
(for 1-dimensional boost expansion $\Theta =1/t$, where $t$ is the proper
time). The same lattice calculations are relevant for the cosmological
quark-hadron transition. Here $\Theta =3H$ where $H$ is the
Hubble-parameter. For $H=1/2t$ and $t$ $\approx 10^{-5}$\thinspace s the
effects of the bulk viscosity in the cosmological quark-hadron transition
are negligible.

But there is another astrophysical setting with a much higher bulk viscosity
where mechanical instability could be relevant, namely neutron stars or
quark stars. For example $P_{\text{eff}}<0$ occurs for a pressure of $%
10^{35} $ dyn/cm$^{2}$, a bulk viscosity of $10^{31}$ g/cm\thinspace s, and
an expansion scalar $\Theta >10^{4}$\thinspace s$^{-1}$, numbers which are
realistic in a compact star setting. In the following the conditions
necessary for cavitation are derived, the relevant adiabatic indices for
quark matter in compact stars are calculated, and the parameter ranges where 
$P_{\text{eff}}<P_{\text{vap}}$ are presented. Mechanical instability turns
out not to occur for nuclear matter, but it can play an important role in
compact stars containing kaon condensation or quark matter. This may have
interesting consequences for compact star phenomenology.

The conditions necessary for mechanical instability in dense matter are best
described in terms of adiabatic indices, defined as $\Gamma \equiv d\ln
P/d\ln n_{B}$, where $P$ is the pressure and $n_{B}$ is the baryon density.
As demonstrated in \cite{Lindblom:2001hd} the bulk viscosity in a dense,
relativistic fluid can be expressed as%
\begin{equation}
\zeta =\frac{P(\Gamma _{\text{Fr}}-\Gamma _{\text{Eq}})\tau }{1+(\omega \tau
)^{2}},
\end{equation}%
where $\omega $ is the frequency of an oscillation, $\tau $ is the
equilibration time-scale for the relevant microscopic processes, and $\Gamma
_{\text{Fr}}$ and $\Gamma _{\text{Eq}}$ are the adiabatic indices for frozen
chemical composition and equilibrium composition respectively. The bulk
viscosity measures the impact of deviation from equilibrium via the
competition between the time-scale characterizing the microscopic response, $%
\tau $, and the variation in fluid density with frequency $\omega $ and it
enters the relativistic hydrodynamical equations in the combination 
\begin{equation}
P_{\text{eff}}=P-\zeta \Theta =P\left( 1-\frac{(\Gamma _{\text{Fr}}-\Gamma _{%
\text{Eq}})\omega \tau }{\left( 1+(\omega \tau )^{2}\right) }\frac{\Theta }{%
\omega }\right) .
\end{equation}%
This equation demonstrates that the effect of bulk viscosity is largest for $%
\omega \tau \approx 1$, and it shows that the condition for cavitation, $P_{%
\text{eff}}<P_{\text{vap}}$, is%
\begin{equation}
\frac{(\Gamma _{\text{Fr}}-\Gamma _{\text{Eq}})\omega \tau }{\left(
1+(\omega \tau )^{2}\right) }\frac{\Theta }{\omega }>1-\frac{P_{\text{vap}}}{%
P}.
\end{equation}%
In particular a necessary condition for cavitation is%
\begin{equation}
\Delta \equiv (\Gamma _{\text{Fr}}-\Gamma _{\text{Eq}})>\frac{2\omega }{%
\Theta }\left( 1-\frac{P_{\text{vap}}}{P}\right) \equiv \Delta _{\text{C}},
\label{necessary}
\end{equation}%
for which $P_{\text{eff}}<P_{\text{vap}}$ in the following range,%
\begin{equation}
\frac{\Delta -\sqrt{\Delta ^{2}-\Delta _{\text{C}}^{2}}}{\Delta _{\text{C}}}%
<\omega \tau <\frac{\Delta +\sqrt{\Delta ^{2}-\Delta _{\text{C}}^{2}}}{%
\Delta _{\text{C}}}.  \label{range}
\end{equation}

\bigskip For a sinusoidal volume oscillation with amplitude $A$ ($%
V=V_{0}(1+A\sin (\omega t))$) and non-relativistic flow velocities, $\Theta
=A\omega \cos (\omega t)/(1+A\sin (\omega t))$, so for not too large
amplitude $\Delta _{\text{C}}\approx 2\left( 1-P_{\text{vap}}/P\right)
/A\cos (\omega t)$. This shows that cavitation is most likely near a phase
boundary ($P\rightarrow P_{\text{vap}}$), that it is facilitated by a large
amplitude, and that the effect of bulk viscosity is most important when $%
\cos (\omega t)\rightarrow 1$, i.e. when the expansion rate is maximal,
whereas the impact on phase-transformation from the change in thermodynamic
pressure is most important for $\sin (\omega t)\rightarrow -1$.

The bulk viscosity of nuclear matter in neutron stars with and without
hyperons has been presented for various equations of state in the literature 
\cite%
{Jones:2001ya,Lindblom:2001hd,Nayyar:2005th,Chatterjee:2007qs,Chatterjee:2007iw}%
. Published numbers are often close to the interesting regime for mechanical
instability, but the necessary condition Eq.\thinspace (\ref{necessary}) may
be hard to fulfill unless the amplitude is very large. For example the
values of $\Gamma _{\text{Fr}}$ and $\Gamma _{\text{Eq}}$ published in
Ref.\thinspace \cite{Haensel:2002qw} differ at most by $0.8$. The highest
published value for $\Delta $ (for neutron stars with kaon condensates \cite%
{Chatterjee:2007qs}) is around $5$. For equations of state that involve
phase transitions, such as neutron stars with a kaon condensed core \cite%
{Kaplan:1986yq}, the relevant vapor pressure for the core is the pressure of
the low-density phase at the core boundary, so $\Delta _{\text{C}%
}\rightarrow 0$ near the boundary, which means that there will be an outer
layer in the core where cavitation triggered by bulk viscosity can take
place for a range of $\omega \tau $ in addition to phase transformation
caused by the oscillation in the thermodynamic pressure.

Bulk viscosities in dense quark matter are known to be even higher than in
nuclear matter for some ranges of parameters \cite%
{Wang:1985tg,Sawyer:1989uy,Madsen:1992sx,Madsen:1998qb,Madsen:1999ci,Alford:2006gy,Sa'd:2006qv,Sa'd:2007ud,Alford:2007rw,Manuel:2007pz,Alford:2008pb}%
, and therefore the most likely setting for cavitation is quark matter in
strange stars if quark matter is absolutely stable or in the inner parts of
neutron stars if quark matter becomes stable only at high pressure. As
demonstrated below, the necessary condition for mechanical instability is
indeed fulfilled in the outer tens of meters of strange stars, and for other
parameters where strange matter is not absolutely stable cavitation can
occur in quark matter cores deeper inside neutron stars.

Quark matter will be treated within the MIT bag model where the pressure is
given by%
\begin{equation}
P=-B+P_{u}(\mu _{u})+P_{d}(\mu _{d})+P_{s}(\mu _{s})+P_{e}(\mu _{e}),
\end{equation}%
where $B$ is the bag constant to be thought of as the energy density of the
confining vacuum, and $P_{i}(\mu _{i})$ are the Fermi-gas contributions from
the constituent particles, which are up, down, and strange quarks as well as
electrons \cite{Witten:1984rs,Farhi:1984qu,Haensel:1986qb,Alcock:1986hz}.
The star is assumed transparent to neutrinos. The temperature is set to zero
for calculations of adiabatic indices (but not for calculations of $\tau $),
which is an excellent approximation after the first few seconds in the life
of a proto-neutron star. The pressure contributions are (for massless up and
down quarks and electrons, and strange quarks with mass $m$ and Fermi
momentum $k_{s}\equiv \left( \mu _{s}^{2}-m^{2}\right) ^{1/2}$)%
\begin{align}
P_{u}(\mu _{u})& =\frac{\mu _{u}^{4}}{4\pi ^{2}},\,P_{d}(\mu _{d})=\frac{\mu
_{d}^{4}}{4\pi ^{2}},\,P_{e}(\mu _{e})=\frac{\mu _{e}^{4}}{12\pi ^{2}}, 
\notag \\
P_{s}(\mu _{s})& = \\
& \frac{1}{4\pi ^{2}}\left[ \mu _{s}k_{s}\left( \mu _{s}^{2}-\frac{5}{2}%
m^{2}\right) +\frac{3}{2}m^{4}\ln \left( \frac{\mu _{s}+k_{s}}{m}\right) %
\right] ,  \notag
\end{align}%
and the corresponding number densities are $n_{i}(\mu _{i})=\partial
P/\partial \mu _{i}$.

The total baryon density $n_{B}=\left[ n_{u}(\mu _{u})+n_{d}(\mu
_{d})+n_{s}(\mu _{s})\right] /3$, and local charge neutrality (which is
obeyed under most conditions except in a mixed quark-hadron phase) gives the
constraint $\frac{2}{3}n_{u}(\mu _{u})-\frac{1}{3}n_{d}(\mu _{d})-\frac{1}{3}%
n_{s}(\mu _{s})-n_{e}(\mu _{e})=0.$

If reactions are very slow compared to the time-scale for density change,
the composition is effectively frozen, and the highest possible value of the
adiabatic index is obtained. To calculate $\Gamma _{\text{Fr}}=d\ln P/d\ln
n_{B}|_{\text{Freeze}}=(n_{B}/P)dP/dn_{B}|_{\text{Freeze}}$ one notes that%
\begin{align}
\frac{dP}{dn_{B}}|_{\text{constraints}}& =\sum_{i=u,d,s,e}\frac{\partial P}{%
\partial \mu _{i}}|_{n_{B}}\frac{\partial \mu _{i}}{\partial n_{B}}|_{\text{%
constraints}}  \label{dPdnconstraints} \\
& =\sum_{i=u,d,s,e}n_{i}\frac{\partial \mu _{i}}{\partial n_{B}}|_{\text{%
constraints}}.
\end{align}
Fixed composition means that the individual densities vary in proportion to
the baryon density, leaving $x_{i}\equiv n_{i}/n_{B}$ constant, so for
example $\partial \mu _{u}/\partial n_{B}=\partial (\pi
^{2}x_{u}n_{B})^{1/3}/\partial n_{B}=\mu _{u}/3n_{B}$, and $\partial \mu
_{s}/\partial n_{B}=k_{s}^{2}/3\mu _{s}n_{B}$. Collecting terms this gives%
\begin{equation}
\Gamma _{\text{Fr}}=\frac{4\left[ P+B+\left( n_{s}k_{s}^{2}/4\mu
_{s}-P_{s}\right) \right] }{3P},
\end{equation}%
reproducing the well-known limit for massless quarks, $\Gamma =4(P+B)/3P$,
which approaches the equally well-known value $\Gamma =4/3$ for a gas of
extremely relativistic particles in the limit where $P\gg B$.

If all particle reactions are very fast compared to the dynamical time-scale
for density change, the composition has time to equilibrate, and the
adiabatic index will take its minimum value, $\Gamma _{\text{Eq}}$. The
values of $x_{i}$ will no longer be constant. Instead chemical equilibrium
conditions are determined by the possible processes, which are%
\begin{align}
s+u& \leftrightarrow u+d \\
d& \leftrightarrow u+e^{-}+\overline{\nu }_{e}  \notag \\
s& \leftrightarrow u+e^{-}+\overline{\nu }_{e}  \notag
\end{align}%
leading to constraints on the chemical potentials%
\begin{equation}
\mu _{s}=\mu _{d}=\mu _{u}+\mu _{e},
\end{equation}%
which together with the definition of baryon density and the condition for
charge neutrality gives 4 constraints on the 5 variables taken to be the
four chemical potentials plus $n_{B}$ or the four particle densities and $%
n_{B}$, which again allows the calculation of all necessary derivatives in
Eq.\thinspace (\ref{dPdnconstraints}) as a function of $n_{B}$. The
calculation involves implicit differentiation of a fairly long expression
for the relation between $n_{u}$ and $n_{B}.$ Therefore only the numerical
results are given here. Again, in the limit of massless quarks, the
expression for $\Gamma _{\text{Eq}}$ converges to $4(P+B)/3P$, which is a
useful consistency check on the results, and also confirms that $\Gamma _{%
\text{Eq}}=\Gamma _{\text{Fr}}$ for massless quarks.

To appreciate the numerical results for $\Delta \equiv (\Gamma _{\text{Fr}%
}-\Gamma _{\text{Eq}})$ shown in Figure 1 as a function of baryon density, $%
n_{B}$, it is important to note that the surface of a stable, self-bound
strange star (possible for $145\,$MeV $<B^{1/4}<165\,$MeV, depending on the
value of $m$ \cite{Witten:1984rs,Farhi:1984qu,Haensel:1986qb,Alcock:1986hz})
is characterized by $P=0$, and that the density increases from zero to well
above nuclear matter density when crossing the surface. For massless quarks
the surface density is exactly $4B$ and contrary to an ordinary neutron star
(where the density, not just the pressure, goes to zero at the stellar
surface), this density increases only by a factor of a few from stellar
surface to stellar center \cite%
{Witten:1984rs,Farhi:1984qu,Haensel:1986qb,Alcock:1986hz}. This unusual
behavior of self-bound matter near the stellar surface is reflected in a
very special behavior of the adiabatic indices. Again for massless quarks
one sees that $\Gamma =4(P+B)/3P$ diverges at the surface, since $%
P\rightarrow 0$, so non-interacting quark matter has the special property
that the equation of state is extremely stiff (high $\Gamma $) at the lowest
(albeit still very high) densities, whereas the equation of state is soft ($%
\Gamma \rightarrow 4/3$) at the highest densities. And for $m>0$ it leaves
room for a significant numerical difference between the adiabatic index
calculated for frozen composition and equilibrium composition respectively
as illustrated in Figure 1. 
\begin{figure}[tbp]
\begin{center}
\includegraphics[
trim=0.000000in 0.000000in 0.000000in 3.157758in,
height=3.4445in,
width=3.5163in
]{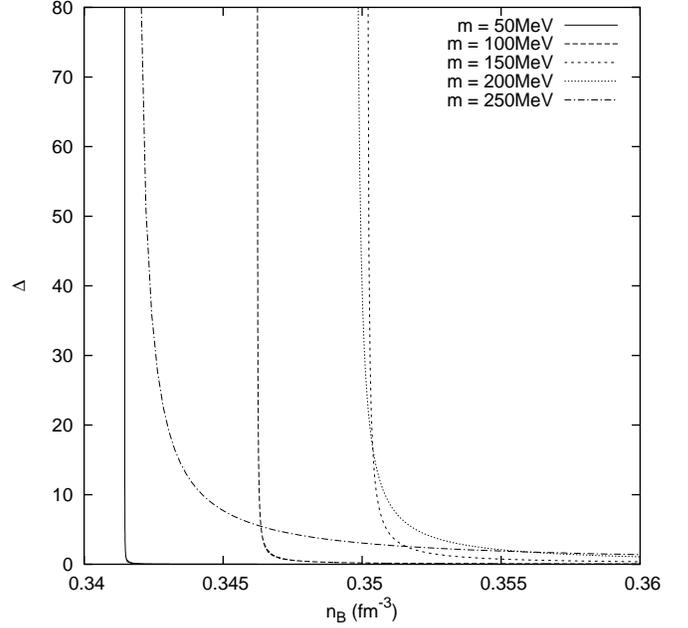}
\end{center}
\caption{Difference in adiabatic indices, $\Delta \equiv \Gamma _{\text{Fr}%
}-\Gamma _{\text{Eq}}$, plotted against baryon number density, $n_{B}$ for
bag constant $B^{1/4}=155\,$MeV and strange quark mass $m$ as indicated. All
curves are diverging near the density where $P=0$ (which varies slightly
with $m$ for fixed $B$). Except for a shift to higher density, similar
curves are obtained for higher values of $B$ where strange matter is
unstable but may form a core in neutron stars.}
\label{Figure 1}
\end{figure}

Figure 1 illustrates that the necessary condition for mechanical
instability, Eq.\thinspace (\ref{necessary}), is obeyed near the surface of
a strange star, where cavitation involves $P_{\text{vap}}=0$ and therefore $%
\Delta _{\text{C}}\approx 2/A\cos (\omega t)$ for sinusoidal perturbations.
For a typical compact star mass of 1.4 solar masses $\Delta >10$ in a
surface layer stretching to a depth of 100 meters and $\Delta >100$ in the
outer few meters. If quark matter is not absolutely stable but found only as
a quark matter core inside a neutron (or so-called hybrid) star, the outer
meters of this core will obey Eq.\thinspace (\ref{necessary}) for high $A$.
In this case the relevant vapor pressure is the pressure of the nuclear
matter phase at the core boundary, $P_{\text{vap}}=P_{\text{ph}}$, so $%
\Delta _{\text{C}}\rightarrow 0$ near the quark core boundary where $%
P\rightarrow P_{\text{ph}}$. If quark matter occurs in a mixed quark-hadron
phase the results above cannot be directly applied, because quark matter in
such a mixed phase does not obey local charge neutrality. However there is
reason to believe that a similar effect will occur there.

When the necessary condition for cavitation is fulfilled, the actual
behavior is determined by $\omega \tau $. The microscopic relaxation time, $%
\tau $, has been calculated for the relevant processes, and can take
essentially all values from $10^{-10}-10$\thinspace s depending on
temperature and strange quark mass in particular \cite%
{Wang:1985tg,Sawyer:1989uy,Madsen:1992sx,Madsen:1998qb,Madsen:1999ci,Alford:2006gy,Sa'd:2006qv,Sa'd:2007ud,Alford:2007rw,Manuel:2007pz,Alford:2008pb}%
. Cavitation is most likely for $\omega \tau \approx 1$ (Eq.\thinspace (\ref%
{range})), and this can very well occur, since many dynamical processes in
compact stars reach such time-scales, including stellar radial oscillations,
r-mode instabilities \cite{Andersson:1997xt}, and tidal deformation during
the final stages of binary compact star inspiral \cite{Bauswein:2008gx}.
Therefore the consequences can be important for detailed studies of these
phenomena and related observables such as the gravitational wave emission
signatures, pulsar glitches and possible triggering of phase transitions
that could be relevant for gamma-ray bursts and other energetic sources.
Also, the detailed break-up process in binary inspiral of strange stars will
decide the mass distribution of fragments in form of strangelets sought for
in cosmic rays \cite{Madsen:2004vw,Bauswein:2008gx,Han:2009sj}.

Quark matter has here been treated within the simplest version of the MIT
bag model without inclusion of gluon-exchange corrections. Such corrections
have been shown (as far as the equation of state is concerned for small
strong coupling) to be equivalent to changing the value of $B$ \cite%
{Farhi:1984qu}, so the qualitative scenario should remain unchanged. Going
beyond the bag model will clearly change the numerical estimates, but the
possibility of having large and even divergent values of the adiabatic
indices, and therefore possibly also large values of $\Delta$, appears
linked to the special properties of self-bound matter rather than to the
specific equation of state. Quark pairing as is expected to occur with color
superconductivity and color-flavor locking in the infinite density limit 
\cite{Alford:2007xm}\ may have only minor effect on the equation of state
and adiabatic indices, but it significantly (often exponentially) decreases
the bulk viscosity via increasing the characteristic microscopic time-scale, 
$\tau$ \cite%
{Madsen:1999ci,Alford:2006gy,Sa'd:2006qv,Sa'd:2007ud,Alford:2007rw,Manuel:2007pz,Alford:2008pb}%
. This would significantly change the mechanical instability window, at
least for full color-flavor locking whereas the change in bulk viscosity is
smaller for phases with partial pairing, such as 2SC. However, pairing is
expected to be most important at the highest densities, and it could well be
that quark matter is unpaired at the lowest densities where the mechanical
instability occurs.

The bulk viscosity in dense quark matter has been shown to make the
effective pressure smaller than the vapor pressure under conditions relevant
for neutron stars or strange stars. This indicates the onset of cavitation
where the quark matter breaks apart into fragments which may survive as
strangelets (if quark matter is stable) or hadronize in a manner comparable
to recent scenarios for ultra-relativistic heavy-ion collisions. At the very
least it points to the need for a more detailed investigation of
instabilities and oscillations in compact stars containing quark matter. In
particular, the cavitation process represents a new mechanism for damping
oscillations, r-modes, and other perturbations in compact stars. Similar
phenomena can occur in the outer parts of kaon condensed cores in neutron
stars (and for other systems with a phase transition separating a core from
the outer parts). Other types of dense nuclear matter do not seem to fulfill
the necessary condition for cavitation, but the numbers are sufficiently
close that also this may deserve further study. The effect of bulk viscosity
is complementary to phase transformations caused by the perturbation in the
thermodynamic pressure itself. The interplay between these contributions and
questions related to the spectrum and growth rates of cavities \cite%
{Torrieri:2008ip} remain to be studied.

\acknowledgments I thank Mark Alford and Krishna Rajagopal for very useful
comments. This work was supported by the Danish Natural Science Research
Council.

\end{document}